\documentclass[journal=jacsat,manuscript=article]{achemso}

\usepackage[version=3]{mhchem} 
\usepackage[utf8]{inputenc}
\usepackage[T1]{fontenc}
\usepackage{xcolor}

\author{Krzysztof Sobucki}
\affiliation
{Institute of Spintronics and Quantum Information, Faculty of Physics, Adam Mickiewicz University, Uniwersytetu Poznańskiego 2, 61-614 Poznań, Poland}
\email{krzsob@amu.edu.pl}

\author{Igor Lyubchanskii}
\affiliation{in association with Adam Mickiewicz University}

\author{Maciej Krawczyk}
\affiliation
{Institute of Spintronics and Quantum Information, Faculty of Physics, Adam Mickiewicz University, Uniwersytetu Poznańskiego 2, 61-614 Poznań, Poland}

\author{Paweł Gruszecki}
\affiliation
{Institute of Spintronics and Quantum Information, Faculty of Physics, Adam Mickiewicz University, Uniwersytetu Poznańskiego 2, 61-614 Poznań, Poland}

\title{Goos-H\"anchen shift of inelastically scattered spin-wave beams and cascade nonlinear excitation of spin-wave leaky modes}

\abbreviations{IR,NMR,UV}

\begin{document}







\begin{abstract}
We show, by means of micromagnetic simulations, inelastic scattering of spin-wave beams on edge-localized spin waves modes. The outcome of the investigated inelastic scattering is creation of new spin-waves beams of frequencies shifted by the edge mode frequencies. We report that inelastically scattered spin-wave beams in both stimulated splitting and confluence processes not only change their direction of propagation, but also undergo lateral shifts along the interface, analogous to the Goos-H\"anchen effect. We report that the lateral shifts of inelastically scattered beams can be much larger than the classical Goos-H\"anchen shifts for reflected spin waves, taking both positive and negative values. In addition, we report the cascade nonlinear excitation of spin-wave leaky modes accompanied by a substantial increase in the value of the lateral shift of the inelastically scattered spin-wave beam in the confluence process. Our results are an important contribution to the understanding of the nonlinear nature of spin waves, which is crucial for spin wave applications. 
\end{abstract}

\section{Introduction}

    Spin waves (SWs), propagating precessional magnetization disturbances, are a promising candidate for information carriers, especially in the context of their applications in beyond-CMOS applications\cite{chumak2022}.
    One of the SWs advantages is their intrinsic nonlinearity, a key element for advanced applications such as neuromorphic computing.
    Another advantage is the possibility of the SW's confinement in a small part of a magnetic material. For instance, SWs may be confined in a nanoscale-wide potential well induced by the static demagnetization field near the film’s edge\cite{lara2013,sebastian2013, hermsdoerfer2009, gruszecki2021}. Such edge-localized SW modes are called edge SWs or edge modes, and usually, their frequencies are lower compared with the SWs propagating outside of the well.     
    Recently, there has been considerable attention devoted to the research on inelastic scattering of SWs on localized modes, mostly to obtain frequency combs\cite{Yao2023, wang2021skyrmion, zhou2021combDW, wang2022vortex}, but also for other applications\cite{gruszecki2022, wang2018dmi}.
    Several spatially localized modes on which propagating SWs are inelastically scattered have been considered. Nonlinear scattering on the skyrmion gyrotropic mode\cite{gareeva2018collective, PAIKARAY2021167900, wang2021skyrmion}, azimuthal SWs in vortex\cite{shi2017transverse, verba2021theory, wang2022vortex,gao2023interplay}, domain wall mode\cite{dadoenkova2019,Wojewoda_neel_domain,zhou2021combDW}, and an SW edge mode \cite{gruszecki2022} were studied. 
    In the last case, the inelastic scattering of an obliquely incident SW beam of frequency $f$ at the edge of a Permalloy (Py) thin film with a propagating SW edge mode of frequency $\nu$ results in two primary three-magnon processes~\cite{gruszecki2022}, i.e., 
    stimulated splitting process~\cite{ordonez2009three, zhang2018, korber2020nonlocal} (SSP) and confluence process~\cite{ordonez2009three, zhang2018} (CP). 
    CP causes two modes at frequencies $f$ and $\nu$ to merge (confluence) into a new SW at frequency $f+\nu$. 
    On the other hand, SSP causes 
    the splitting of the mode at frequency $f$ into two modes at frequencies $f-\nu$ and $\nu$  with the assistance of the mode at frequency $\nu$ stimulates this process and increases the intensity of new waves created in SSP over CP. It has been shown that SSP could be used to realize the magnon-based transistor \cite{ge2024nanoscaled}, while both SSP and CP could be used for demultiplexing \cite{gruszecki2022}. In the second example, 
    the effect used is that the scattered SW beams created in nonlinear processes propagate under different angles compared to the reflected SW beam and the angle of propagation depends on the edge mode frequency $\nu$. The mechanism of this phenomenon was explained by employing the isofrequency contours analysis and conservation of the tangential component of beam wavevectors~\cite{lock2008properties, gruszecki2017graded}.
    
    The reflection of the wave beam from the edge of the material is in some cases associated with the Goose-H\"anchen (GH) effect. This effect was first observed and described in optics~\cite{goos1947,Renard64,Snyder76} and manifests itself as a spatial shift of a totally internally reflected light beam along the interface. The origin of this effect is the phase acquisition of the waves during reflection. In optical experiments, the GH shift ranges from a few nanometers (fractions of light wavelengths) up to few micrometers~\cite{Wild_1982, Wang2008_GHE_light_beam, Wang2013_GHE, wang2021targeted}. 
    As GH effect stems from the wave nature, the analogous effects of the spatial shift of different types of waves can be also observed. For instance GH effect was confirmed for plasmons~\cite{Yin2006, Parks:15, Zeng20_GHE_plasmons} and SWs \cite{dadoenkova2012huge, gruszecki2014,gruszecki2017goos,stigloher2018,wang2019GH,laliena2022}. Besides, it has been theoretically predicted that the lateral shift of an optical beam undergoing Brillouin light scattering by phonons \cite{jiang2013electrically, villegas2017goos, dadoenkova2018, zhou2019precise, Zhang:21} can occur. On top of that, it was shown that GH-like effect also takes place during the inelastic scattering of electromagnetic waves on magnons \cite{rajeswari2014intensities, Santos2018, dadoenkova2019, dadoenkova2022}. However, up to date, there is no report on the GH effect of the inelastically scattered SW beams.
    
    In this paper, we report the GH shift for the SW beam that is inelastically scattered on the propagating edge mode. We numerically demonstrate this effect in in-plane magnetized thin Py film as a result of the SSP and CP nonlinear processes taking place at the very edge of the film. We show that the edge mode frequency and wavevector, as well as the angle of propagation of the incident SW beam, affect the scattered beams. Unexpectedly, we found a significant enhancement of the GH shift of the SW beam generated in the CP process at certain frequencies of the edge mode. We show that this effect is related to the cascade of three nonlinear processes involving the edge mode and generated high-frequency edge and leaky waves propagating into the film. 
    These findings demonstrate new effects in magnonics and provide a basis for new methods of SW beam control for practical use in magnonic technology.

    \section*{\label{sec:res} Results}

\begin{figure}[!h]
        \centering
        \includegraphics[width=0.9\columnwidth]{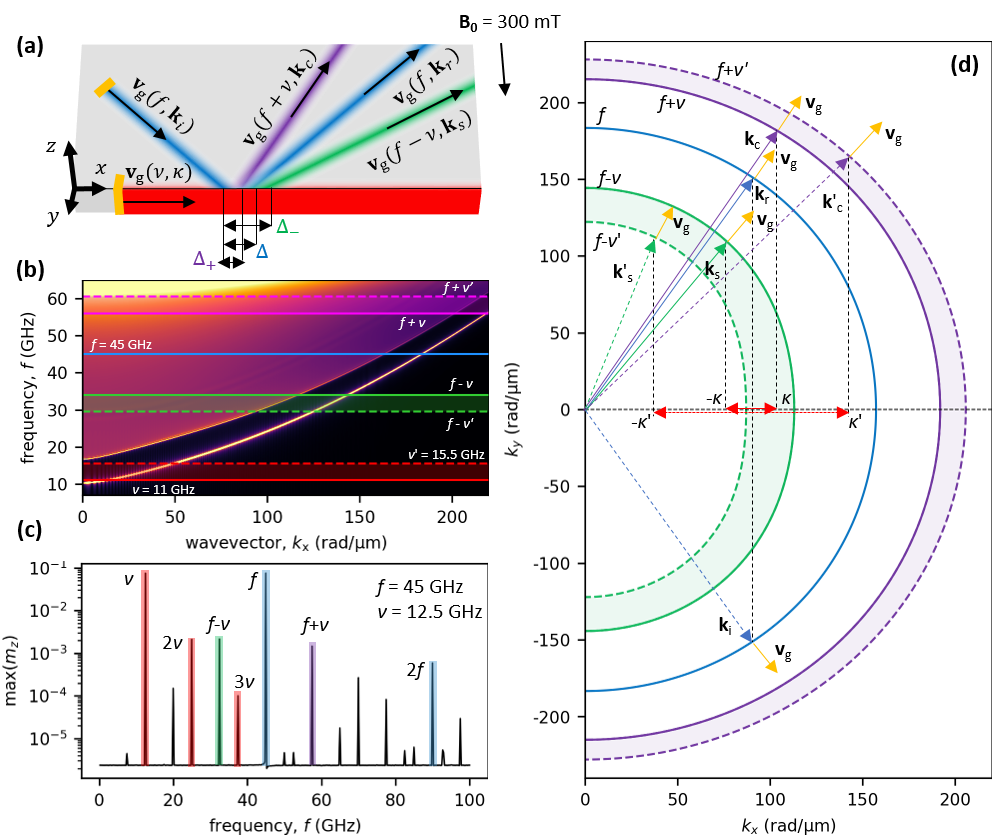}      
        \caption{ \protect\footnotesize \textbf{System's geometry and dispersion relation with isofrequency contours construction.}
        \textbf{a} Geometry of the investigated system, i.e., Py layer with thickness $10$~nm placed in a uniform external magnetic field $\mu_0 H_0=300$~mT along the $-y$~axis with a schematic representation of SWs propagating at different frequencies (not a result of micromagnetic simulations). The blue color represents incident and elastically reflected beams. The purple and green colors represent inelastically scattered beams as a result of CP and SSP, respectively. The red color depicts the edge mode. The black arrows denote the directions of the group velocities $\mathbf{v}_\mathrm{g}$ associated with the SWs propagating in the system at different frequencies.
        The orange lines represent the antennas used to excite the incident SW beam and the edge modes. $\Delta$, $\Delta_{-}$, and $\Delta_{+}$ represent the lateral displacement between the incident beam spot and the beams from reflection, SSP, and CP, respectively.
        \textbf{b} Dispersion relation depending on the tangential component of the wave vector to the interface calculated by means of micromagnetic simulations.
        The red horizontal lines mark the frequency range of the edge modes used in simulations.
        \textbf{c} Spectrum calculated for the system response to scattering of the incident SW beam at a frequency $f=45$~GHz on the edge mode at a frequency $\nu=12.5$~GHz. The blue highlighted peaks correspond to the incident SW beam and its second harmonic. The red peaks represent the edge mode and its harmonics. SSP and CP are highlighted in green and purple, respectively. These frequencies are also marked with horizontal lines in (b).
        \textbf{d} Isofrequency construction illustrating the principles of selecting wave vectors of inelastically scattered SW beams on edge mode propagating to the right, i.e., with $\kappa > 0$. 
        The blue, green, and purple curves represent isofrequency contours for SWs at frequencies of the incident SW beam $f=45$ GHz, reduced in frequency by the edge mode frequency ($f-\nu$ or $f-\nu\prime$) corresponding to SSP, and increased in frequency by the edge mode frequency ($f+\nu$ or $f+\nu\prime$) corresponding to CP. The green and purple isofrequency contours are plotted for two edge mode frequencies, i.e. solid lines for $\nu=11$ and dashed lines for $\nu\prime=15.5$ GHz.
        We keep the same color code as in the previous figures to represent the processes taking place. The red, green, blue, and purple arrows represent successively the wavevectors of the edge SWs, the incident SW beam, and the inelastically scattered SW beams resulting from SSP and CP. 
        In addition, we mark the group velocities of the SW beams with yellow vectors normal to the curvatures of the contours. 
        }
        \label{fig:geo}
    \end{figure}

    We analyze the dynamics of SWs in a 10 nm thick Py layer by means of micromagnetic simulations performed in the Mumax3 environment \cite{vansteenkiste2014design}. We use typical material parameters of Py ($\mathrm{Ni}_\mathrm{80}\mathrm{Fe}_\mathrm{20}$), namely $M_\mathrm{S} = 800$~kA/m, $A_\mathrm{{ex}}=13$~pJ/m, but with reduced damping parameter $\alpha = 0.0001$ for easier analysis of SW propagation in the far field. The layer is semi-infinite, i.e., it has only one sharp edge and is nominally infinite in all other in-plane directions, Fig \ref{fig:geo}a. The external uniform magnetic field $B_0 = \mathrm{\mu_0}H_0=300$~mT is applied perpendicular to the layer's edge, $\textbf{H}_0 = H_0\widehat{\textbf{y}}$, which induces a demagnetizing field on the edge that locally lowers the effective static field. This non-uniformity serves as a potential well in which the SW-localized mode can be confined\cite{Bailleul_2001,McMichael2006,gruszecki2021}. Accordingly, the dispersion relation shown in Fig.~\ref{fig:geo}b consists of a part representing a continuum of SWs freely propagating far from the interface and a distinct band representing edge mode with frequencies downshifted with respect to the dispersion of SWs propagating far from the edge.
    In a low-frequency range, namely, from $11$~GHz to $15.5$~GHz, there is a gap between the edge mode (at wavenumber $k_x=0$) and the SW continuum. This implies that in this frequency range, the propagating SWs are confined to the edge of the system. This situation provides us with a straightforward way to excite only edge modes and to study the inelastic scattering of higher frequency SWs from the thin film on lower frequency modes confined to the film edge.

    We study the process of inelastic scattering of a SW beam of frequency $45$~GHz (corresponding to wavelength of $35.7$~nm), full width at half maximum $756$~nm, incident obliquely on an edge where an edge mode  of $\nu$ frequency is localized.
    Throughout the paper, the frequency of SW edge mode is kept below the bottom of the SW continuum, $15.5$~GHz, so the edge mode cannot leak the energy to the bulk of Py layer. 
    We place two antennas, which emit local oscillating magnetic fields to excite SWs (see yellow stripes in Fig.~\ref{fig:geo}a). 
    The first antenna is placed at the very edge of the system and is responsible for exciting edge mode with frequency $\nu$. 
    The second antenna is placed about 3.84~$\mathrm{\mu m}$ from the edge and excites SW beam with frequency $f=45$~GHz, and corresponding wavevector.
    The second antenna excites unidirectionally propagating SW beam towards the edge\cite{whitehead2019graded, sobucki23nano} at a specific angle of incidence (angle between the wavevector of incident SW beam and normal to the interface, i.e., the $y$-axis).
    More details are given in the Methods section.

    In the numerical simulations we change three parameters, the angle of SW beam incidence $\theta$ and edge mode frequency (which also changes the wavenumber of the edge mode $\kappa$), and the sign of $\kappa$.
    The angle of incidence is controlled by changing the angle of antenna's rotation with respect to the edge of the film. 
    We use the following set of angles $\theta=\{30^\circ, 35^\circ, 40^\circ, 45^\circ\}$. We choose the sign of the edge mode wavevector by changing the position of the edge antenna with respect to the incident beam spot at the interface, i.e., the area where the incident SW beam reaches the edge of the system. If the edge antenna is to the left of the incident SW beam spot, the edge mode wavevector $\kappa$ is positive.  If the antenna is to the right, $\kappa$ is negative.
    For each configuration with chosen $\theta$ and sign of $\kappa$ we run a series of simulations with different edge mode frequencies. These frequencies are in the range of $\nu \in \langle 11, 15.5 \rangle$~GHz.

    Fig. \ref{fig:geo}c shows the spectrum obtained in the steady state of the system in the case of $f=45$~GHz, $\nu=12.5$~GHz, $\theta=30^{\circ}$, which was calculated by averaging over space the absolute value of the Fourier transform from the time to frequency domain calculated for each unit cell independently.
    Two peaks highlighted in blue correspond to the frequency of SW beam $f$ and its second harmonics. 
    Several peaks highlighted in red represent the edge mode $\nu$ and its higher harmonics \cite{gruszecki2021local}. 
    Two peaks of the main interest are marked with green and purple colors. 
    These correspond to the frequencies of SSP $f-\nu$ and CP $f+\nu$ respectively, what confirms the existence of these phenomena in the studied system. We will focus on these effects throughout the paper. 
    Apart from the mentioned peaks, we also see several other peaks at frequencies corresponding to higher-order nonlinear processes (e.g. $f-2\nu=20$ GHz). These are beyond the scope of this paper and will be omitted in further analysis. 

    \begin{figure}[!h]
        \centering
        \includegraphics[width=0.95\columnwidth]{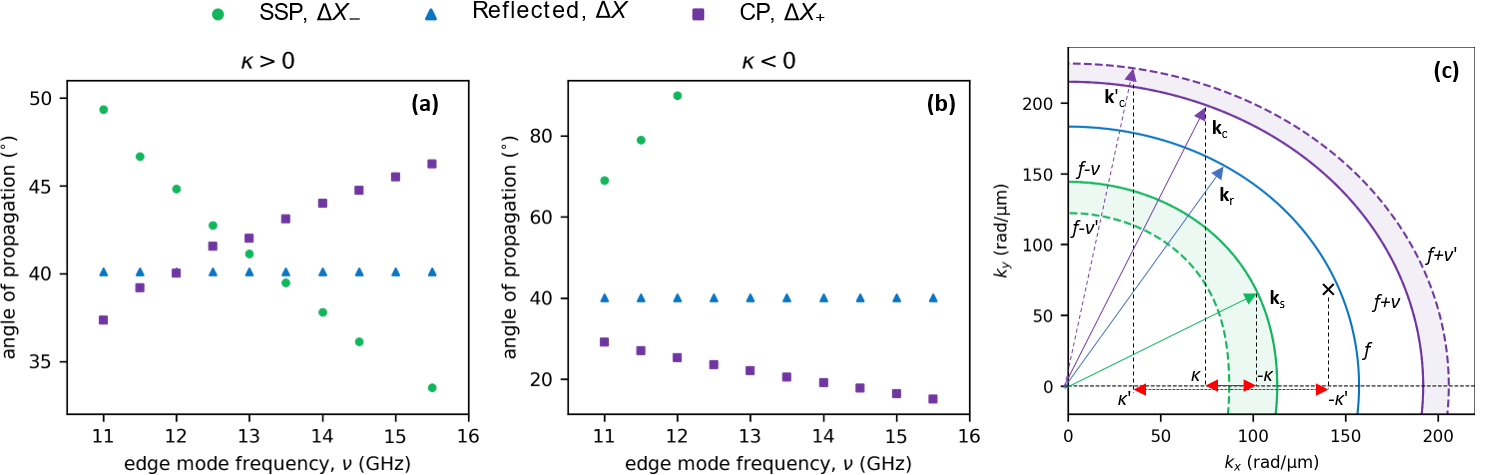}      
        \caption{
        \textbf{Dependencies showing the propagation angle of SW beams.} 
        \textbf{a,b} Angles of propagation of SW beams as a function of edge mode frequency $\nu$, blue triangles represent the reflected beam, green circles display SW beam created in SSP, and purple squares represent the SW beam created in CP.  The results were obtained for the angle of incidence $\theta=30^{\circ}$ and for both, the edge modes with (a) $\kappa > 0$ and (b) $\kappa < 0$. \textbf{c} Isofrequency construction in the case of $\kappa < 0$. Here, for $-|\kappa^\prime|$. There is no geometrical solution for tangential wavevector $k_{r,x} - (-|\kappa^\prime|)$, which would correspond to SWs scattered on higher frequency edge mode. Thus in the case of $\kappa < 0$ there are only a few allowed solutions for SSP, as shown in (b)}. 
        \label{fig:angle}
    \end{figure}

    For edge modes propagating to the right (i.e., $\kappa > 0$), the angles of propagation of the inelastically scattered beams incident at $\theta=30^{\circ}$ change monotonically with the change of $\nu$, see Fig~\ref{fig:angle}a.  
    The angle of propagation of the beam generated in SSP decreases linearly with increasing $\nu$, and changes by $17^{\circ}$ in the examined range of $\nu$. 
    The changes for the beam in CP are in the opposite direction to SSP. Here the angle increases with increasing $\nu$, and in the investigated range of $\nu$ the angle changes by $10^{\circ}$. 
    The results for other angles of propagation of the incident SW beam are qualitatively the same as described in this section.
    These changes in the angle of the propagation of inelastically scattered beams can be explained using isofrequency contour analysis\cite{gruszecki2022}. 
    We present the inelastic scattering of SW beams on edge mode with $\kappa > 0$ as an isofrequency problem in Fig~\ref{fig:geo}d. 
    The blue curve represents all available solutions for the frequency $f$ and $k_x>0$. The blue dashed arrow represents the wavevector associated with the incident SW beam ($\mathbf{k}_\mathrm{i}$).
    The solid blue arrow, marked on the opposite quadrant of space $(k_x, k_y)$ and plotted according to the  conservation of wavevector tangential component ($k_x$) rule (i.e., Snells law), represents the wavevector associated with the reflected SW beam ($\mathbf{k}_r$), which must be located on the isofrequency contour for frequency $f$.
    The beam propagation directions are related to the directions of the group velocities $\mathbf{v}_\mathrm{g}(f,\mathbf{k})$ (marked with orange arrows), which are perpendicular to the constant frequency contour for the corresponding wavevectors.
    Taking into account the conservation of energy ($f^\prime=f + \nu$ for CP and $f^\prime=f - \nu$ for SSP) and the conservation of the tangential component of the wave vector to the interface ($k_{c,x}=k_x + \kappa$ for CP and $k_{s,x}=k_x - \kappa$), a similar construction can be made for inelastically scattered SWs. 
    These contours, for SSP (green curves) and CP (purple curves), are marked for two examples of edge mode frequencies $\nu=11$ GHz and $\nu^\prime=15.5$ GHz, solid and dashed lines respectively.
    Observing the orange arrows corresponding to the group velocity directions, it can be seen that for the lower edge mode frequency, the propagation angle relative to the normal to the interface ($y$-axis) is smaller, which is consistent with the simulation results shown in Fig.~\ref{fig:angle}a.
    A similar construction is done for SSP, and
    , but for the isofrequency contour plotted for $f-\nu$ and for the $x$-component of the wavevector equal to $k_{s,x}=k_x - \kappa$. Therefore, 
    we observe that for the lower value of the edge SWs frequency, the angle of propagation of the inelastically beam increases, which is also in agreement with Fig.~\ref{fig:angle}a.

    In the case of the edge mode propagating leftwards (i.e. $\kappa < 0$, Fig~\ref{fig:angle}b) there are only a few SSP allowed solutions for low frequencies $\nu$. As shown in the isofrequency contour construction in Fig~\ref{fig:angle}c for $k_{s,x}=k_x+|\kappa|$ there are only a couple of solutions for low $\kappa$ vectors that correspond to $f - \nu$ frequencies. These few allowed solutions are shown in Fig~\ref{fig:angle}b in the case $\theta=30^{\circ}$. The angles of propagation of these beams are above $70^\circ$, i.e., their amplitude distribution overlaps with the interface, and therefore it is difficult to derive their trajectories and spatial shifts at the interface. For bigger $\theta$ no SW beams corresponding to SSP are observed.
    The angle of propagation of the beam created in CP decreases with the increase of the edge mode frequency which agrees with the analysis of the changes of the group velocity directions shown in Fig~\ref{fig:angle}c. 
    In the investigated range of the edge mode frequencies, the angle of propagation of the beam created in CP can change by $15^{\circ}$.

    \begin{figure}[!h]
        \centering
        \includegraphics[width=0.5\columnwidth]{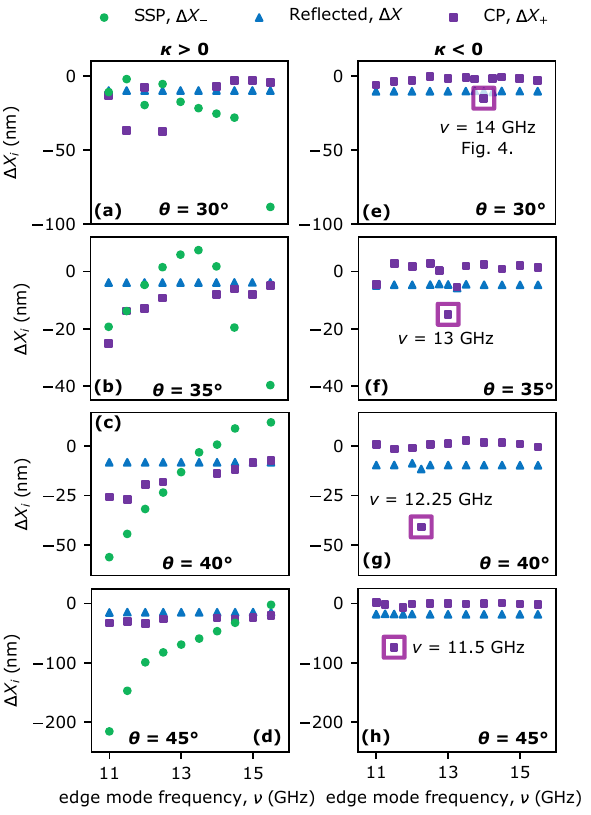} 
        \caption{
        \textbf{Dependence of the lateral shifts along the interface of reflected and inelastically scattered beams on the angle of incidence ($\theta$) and edge SWs frequency ($\nu$) for $\kappa>0$ and $\kappa<0$.}
        The spatial shifts of the elastically reflected beam $\Delta \it{X}$ (blue triangles) and inelastically scattered beams created in SSP and CP $\Delta \it{X}_{-}$, $\Delta \it{X}_{+}$ (green circles and purple squares, respectively) are shown. The left column, (a-d), shows the results of inelastic scattering on edge modes with a positive wavevector, and the right column, (e-h) with a negative wavevector. Additionally, in (e-h) we marked special cases of the results with the big violet squares. 
        }
        \label{fig:ghe}
    \end{figure}

    A detailed analysis of the rays of inelastically scattered SW beams shows that the scattered
    beams are laterally shifted along the interface relative to the incident SW beam spot, see different cases shown in Fig.~\ref{fig:ghe}. This effect is analogous to the GH effect, which appears as a lateral shift of the elastically reflected wave beam with respect to the incident wave beam for electromagnetic waves \cite{goos1947,Renard64,Snyder76} and SWs \cite{dadoenkova2012huge, gruszecki2014,stigloher2018}. 
    The results of simulations with edge mode of wavevector $\kappa > 0$  and for $\kappa < 0$ for different incident beam angles are displayed in Figs~\ref{fig:ghe}(a-d), and Figs~\ref{fig:ghe}(e-h), respectively. We mark spatial shift $\Delta \it{X}$ of the elastically reflected beam with blue triangles, and the inelastically scattered beams with green circles, and purple squares for SSP $\Delta X_\mathrm{-}$ and CP $\Delta_\mathrm{+}$, respectively. 
    The value of the spatial shift for the elastically reflected SW beam does not depend on the frequency of the edge mode. We report the smallest $\Delta \it{X}= -3.8$~nm for the angle of incidence $35^\circ$ and the biggest $-14.5$~nm at $45^\circ$ with $\kappa > 0$. Other works report significant values of GH-shift for special cases \cite{dadoenkova2012huge,zhen2020giant} in strong magnetic fields and with incident SWs propagating at grazing angles to the interface. In a paper with the conditions similar to presented in this work the scope of GH shift for SWs was reported to be in range up to $40$~nm in a system with varying anisotropy and SW wavelength of $60$~nm\cite{gruszecki2017goos}.

    The scope of spatial shifts of the beams created in SSP (green circles) scattered on the edge mode with $\kappa>0$ depends both on the angle of SW incidence beam and the edge mode frequency. At $\theta=30^\circ$ the values of $\Delta \it{X}_{-}$ decrease with increasing $\nu$  almost linearly as shown in Fig~\ref{fig:ghe}(a). For $\theta=35^\circ$ $\Delta \it{X}_{-}(\nu)$ dependency resembles a negative quadratic function with a global maximum in the middle of edge mode frequency range, Fig~\ref{fig:ghe}(b).  In the cases of $\theta=40^\circ$ and $\theta=45^\circ$ $\Delta \it{X}_{-}$, Figs~\ref{fig:ghe}(c,d), increases monotonically with increasing edge mode frequency. The values of $\Delta \it{X}_{-}$ in most of the cases are negative and in the range of tens of nanometers, thus in comparison with the incident beam width the spatial shifts are relatively small. The wavelengths of the scattered beams created in SSP slightly vary with different angles and edge mode frequencies but in the simulations in range between $48$~nm and $58$~nm. Thus reported spatial shifts are comparable or smaller than scattered SWs wavelengths. Exceptions to this rule are the results for $\theta=45^\circ$ and low edge mode frequencies which are in the range of hundreds of nanometers, e.g. for $\nu=11$~GHz $\Delta \it{X}_{-}=-215$~nm. 
    In the results presented in Figs~\ref{fig:ghe}(a-d) we omit edge mode frequency $\nu=15$~GHz as the frequency resulting in SSP is $f-\nu=30$~GHz, which is the second harmonic of edge mode excitation that overlap with the bulk SW spectra thus interfering with the derivation of SW beam trajectory.
    
    The dependencies of spatial shifts of beams created in CP (violet squares) in $\kappa>0$ are straightforward. For all incidence angles $\theta$ $\Delta \it{X}_{+}(\nu)$ increases with the edge mode frequency increase. The scope of $\Delta \it{X}_{+}$ is smaller in comparison to the spatial shifts calculated for beams created in SSP, and are in the range of tens of nanometers thus, they are small in comparison to  scattered SWs wavelengths which for CP are in range of $28$~nm and $30$~nm. Additionally, in the Figs~\ref{fig:ghe}(a-d) shifts for frequencies $\nu=13$~GHz and $\nu=13.5$~GHz are not displayed because for these frequencies the amplitudes of the scattered beams are very low, and derivation of their trajectories is unreliable. 

    Figs~\ref{fig:ghe}(e-h) show the results of spatial shift derivation of SW beams scattered on edge mode with $\kappa<0$. In most of the cases the SSP does not occur, as was explained before, and in the few cases that this process does occur the scattered beams are almost grazing the edge thus proper derivation of their spatial shifts is inaccurate. For that reason in the case of  $\kappa<0$ we only present spatial shifts of beams created in CP. In all simulations with different angles of incidence, the derived spatial shifts of the scattered beams in CP are extremely small, in the range of a few nanometers. However,  we can find a certain value of the edge mode frequency for which a significant enhancement of the GH-like shift value appears, see violet squares in Figs~\ref{fig:ghe}(e-h). The origin of this phenomenon will be explained in the following paragraphs.

    \begin{figure}[!h]
        \centering
        \includegraphics[width=0.9\columnwidth]{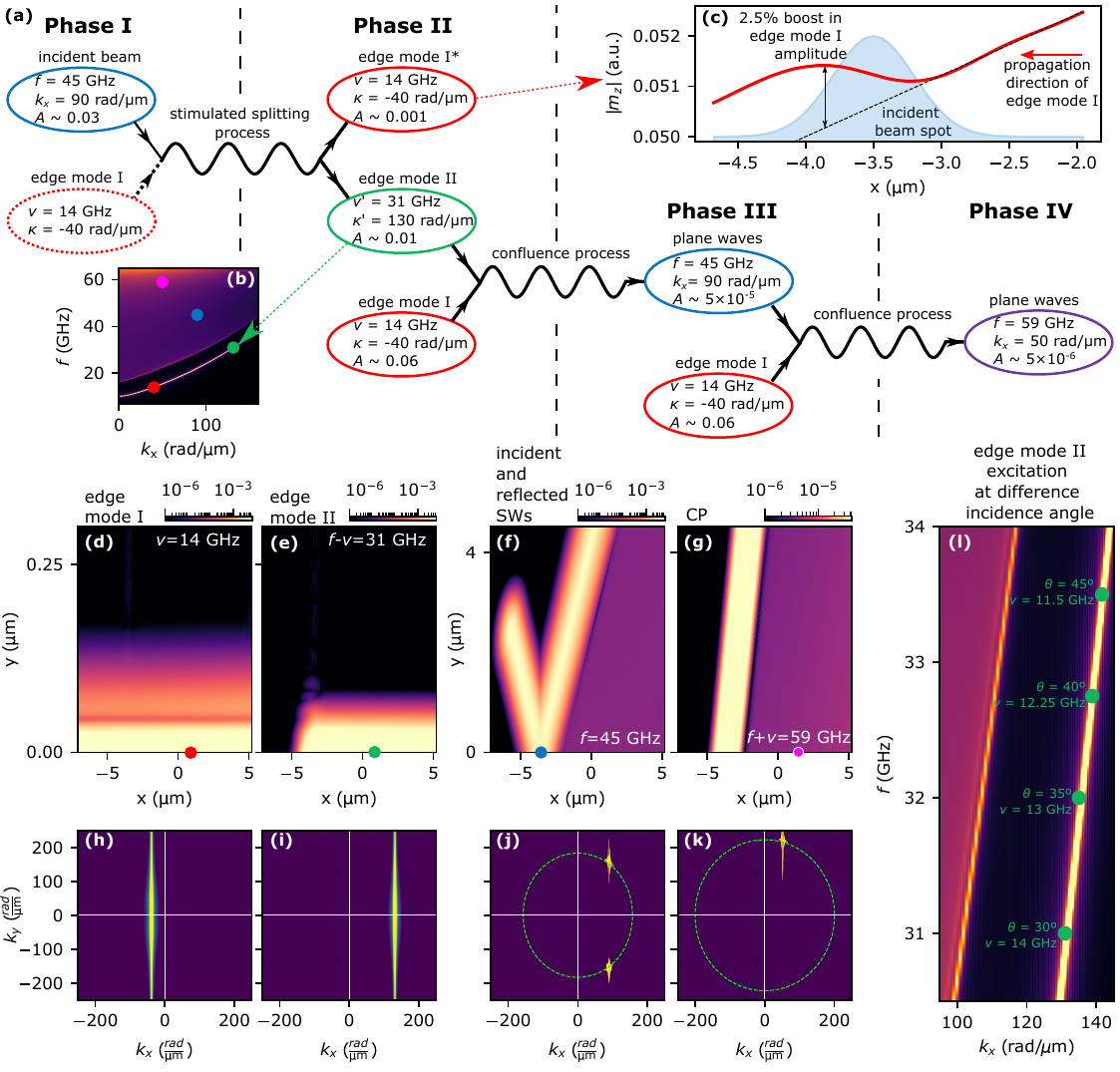}     
        \caption{
        \textbf{Cascade nonlinear excitation of SW leaky modes.}
       (a) Scheme of nonlinear cascade process explaining the excitation of edge and plane SW in the system with incidence SW beam of frequency $f=45$~GHz propagating under angle $\theta=30^{\circ}$. The ovals symbolize SW modes, modes of the same characteristics are marked with the same colors. The proposed cascade process is divided into four phases connected by SSP and two CPs. (b) Dispersion relation of the system with the marked modes shown in (a) (the color of the dots corresponds to the color of the ellipses in (a)). (c) Amplitude $|m_z|$ of the edge mode I (red, solid line) at the film edge in the vicinity of the incidence SW beam spot. The black, dashed line is the extrapolation of  $|m_z|$  simulated dependency before the incidence spot, i.e., for $x>-3 \mu m$. 
       (d-g) Space distributions of $|m_z|$ at frequencies $\nu=14$~GHz, $f-\nu=31$~GHz, $f=45$~GHz, $f+\nu=59$~GHz, respectively. The dots indicate spots where SW modes amplitudes, $A$ indicated in (a) were taken. 
       (h-k) Results of the two-dimensional Fourier transform from the space to wavevector domain of SW amplitude distribution shown in (d-g). Figures (j,k) additionally contain isofrequency contours that correspond to frequencies of SWs presented in these pictures. (l) Investigated system's dispersion relation with marked frequencies for which excitation of SW plane waves occurs at different angles of incidence.
     }
        \label{fig:cascade}
    \end{figure}


    Let us first focus on the case shown in Fig.~\ref{fig:ghe}(e), i.e., for the angle of incidence 30$^\circ$ and the frequency of the edge mode $\nu=14$ GHz.
    The spatial distributions of the SW amplitude at frequencies $\nu=14$ GHz, $f=45 GHz$ and $f+\nu=59$ GHz are shown in Fig.~\ref{fig:cascade}(d), (f) and (g), respectively. Surprisingly, in addition to the beams (incident and reflected at 45 GHz, and scattered at 59 GHz), we can also see a brightened region representing the plane waves with the same wave vector as the SWs from the beam.
    This effect can be explained as a result of a cascade nonlinear excitation of plane waves at the film's edge.
    Fig.~\ref{fig:cascade}(a) shows a scheme of the proposed nonlinear cascade process in the case of an incident SW beam propagating under angle $\theta=30^{\circ}$ and scattered on a propagating edge mode of frequency $\nu=14$~GHz. The first phase of this process is the SSP of the incident SW beam, marked by the blue oval, on the edge mode I, dashed red oval. As described before, there is no allowed solution to this process in the bulk of the Py layer. However, in this particular case, the result of the SSP is the generation of SWs with frequency $f-\nu=31$~GHz and wave vector $k_x-|-\kappa|=130$~rad/$\mathrm{\mu}$m which coincide with one of the allowed edge states in the system, as shown in Fig.~\ref{fig:cascade}(b), later in this paper we will call this mode, edge mode II. Fig.~\ref{fig:cascade}(e) shows the spatial distribution of the SW amplitude of this mode, confirming its existence.
    According to the conservation of energy and momentum laws, this SSP must also generate SWs with frequency $\nu=14$~GHz and wave vector $|\kappa|=-40$~rad/$\mathrm{\mu}$m corresponding to the primary excited edge mode (edge mode I, marked in red with an asterisk). 
    The analysis of this edge mode I amplitude presented in Fig.~\ref{fig:cascade}(c) shows a $2.5$\%  boost just behind the spot where the incident beam reaches the edge. It indicates the creation of new edge SWs that propagate along the antenna-excited edge mode I, and can be considered as an amplification of the propagating edge mode\cite{gruszecki2022}. 
    In Fig.~\ref{fig:cascade}(e), where edge mode II is presented, there is a distinctive gap left to the point of SW beam incidence ($x=-3.7$~/$\mathrm{\mu}$m), which indicates that edge mode II is created at the SW beam incidence spot and propagates in opposite direction to edge mode I, Fig.~\ref{fig:cascade}(d). It is evident by calculating Fourier transform from the space to wavevector domain, as presented in Figs~\ref{fig:cascade}(h,i). Both of the edge modes have wavevectors of opposite signs and their numerical values agree with the analytically derived values shown in scheme Fig.~\ref{fig:cascade}(a). 

    The excited edge mode II propagating along the edge interacts with the edge mode I excited directly by the antenna, Fig.~\ref{fig:cascade}(a), Phase II. It leads to CP occurring along edge's length past the point of SW beam incidence spot. The outcome of this process is the creation of a new SW at the edge with $f=45$~GHz and wavevector $k_x=90$~rad/$\mathrm{\mu}$m, which corresponds to the incident SW beam. 
    Taking into consideration the isofrequency contours, newly-created SWs at the edge also have to gain a wavevector component perpendicular to the edge, since there is no such solution for pure edge mode. For that reason, these SW leaks the energy from the edge and propagate into the bulk of the system with their wavefronts parallel to the reflected SW beam. 
    It is shown in Fig.~\ref{fig:cascade}(f), where an area of nonzero SW amplitude is evident right to the reflected beam. Additionally, the spatial Fourier transform of that distribution, shown in Fig.~\ref{fig:cascade}(j), consists of only two distinctive peaks that overlap with isofrequency at $f=45$~GHz and correspond to the incident and reflected SWs. Thus, no SWs with new wavevectors at this frequency are excited in the system. 
    Since we are dealing with a nonlinear cascade process in which the interaction of two edge modes of different frequencies ultimately leads to energy leakage from the edge, this process is a nonlinear analog of a leaky mode excitation, which we have already reported\cite{sobucki23nano}. 
    Therefore, we refer to this phenomenon as nonlinear cascade leaky mode excitation.

    Furthermore, this newly created SW at the edge interacts with the antenna-excited edge mode I, Phase III. This interaction is yet another CP occurring at the system's edge right to the incidence spot. In this CP new SW plane waves are excited, which propagate parallel to the scattered SW beam. We show this wave in Fig.~\ref{fig:cascade}(g) where the distribution of $|m_z|$ at frequency $f=59$~GHz has nonzero amplitude only to the right of the scattered SW beam created in CP, i.e., in the direction of propagating edge mode II. Fig.~\ref{fig:cascade}(k), presenting the space-domain Fourier transform of the $|m_z|$  distribution. There is only one peak that coincides with the isofrequency contour of $f=59$~GHz confirming that the new SW plane wave has the same wavevector as the scattered SW beam. 
    
    In the scheme in Fig.~\ref{fig:cascade}(a) we wrote down the amplitudes ($A$) of all SWs types considered in the described cascade process. It points out that the amplitudes of created SWs in each phase are proportional to the product of the amplitudes in the previous phase. In Figs.~\ref{fig:cascade}(d-g) we mark positions where the $|m_z|$ amplitude values have been taken from the simulation results with circles, which colors correspond to the SWs modes shown in the Fig.~\ref{fig:cascade}(a). The comparison between the incident beam and CP plane waves' amplitudes shows a decrease in the range of $10^{4}$ factor over the course of three process phases. Despite such a minuscule magnitude of new SWs they seem to have a noticeable impact on the lateral shift of scattered beams as shown in Figs~\ref{fig:ghe}(e-h). Indeed, closer analysis of Figs~\ref{fig:cascade}(f,g) shows that there is a narrow, yet distinctive drop in SW amplitude between the beams and plane waves. Such a drop in amplitude is a result of the destructive interference of these two kinds of SWs, thus the newly excited SWs at the edge have to be significantly shifted in phase. 
   
    We also confirmed numerically that the nonzero amplitude background behind the scattered SW beams does not cause a false illusion of the beam shift. As it will be explained in the last paragraph of this section in more detail, the amplitudes of the beams were derived in the data post-processing by fitting Gaussian curves to the amplitude cutlines away from the system's edge. 
    To confirm that adding a small-amplitude background to only one side of the Gaussian curve does not contribute the most to the calculated beam shift, we set up a numerical test. Namely, we added to a Gauss curve (with dimensions corresponding to the simulation results) a Heaviside step function with a height smaller by two orders of magnitude in comparison to the maximal amplitude of the Gauss curve. Then we run the same post-processing for this data as for simulated scattered beams. In the case of a numerically plotted Gauss curve, the addition of amplitude background alters the derived lateral shift by only a few nanometers. In the simulations, the scattered beams undergo lateral shifts up to tens of nanometers. These calculations confirm that adding a small-amplitude background does not change substantially the obtained values of lateral shifts of simulated SW beams.

    In Fig.~\ref{fig:cascade}(l) we show a part of system's dispersion relation where the parameters of edge modes II are marked for all simulated angles of SW beam incidence. It is evident that in the investigated cases of SW beams scattered on the edge mode I with $\kappa<0$ only a very limited number of $(\nu,\kappa)$ and $(f,k_x)$ combinations will yield excitation of new edge modes. The reason for this is the relatively small width of the edge mode band in dispersion relation which allows only a narrow range of SWs to be excited at the edge. For that reason the enhancement of the GH shifts in Figs~\ref{fig:ghe}(e-h) exists only at a narrow range of edge mode I frequencies.

    \section*{Conclusions}

    In summary, 

    showed that the inelastically scattered beams undergo spatial shifts at the interface, what we interpret as an analog of the GH effect.  However, we found that for higher incidence angles and low-frequency edge modes with positive wavevectors the spatial shifts of the beams created in SSP become comparable with beams' width. Moreover, we found a peculiar phenomenon at certain edge mode frequencies and $\kappa <0$. In these cases, we observe excitations of new higher-frequency edge modes in the system when SSP results in additional excitation of already propagating edge modes. Interestingly, this excitation leads to a nonlinear cascade process consisting of two additional confluence processes. These confluence processes create SWs with parameters of the reflected and scattered SW beams in the form of plane waves that propagate parallel to the corresponding SW beams. The proposed cascade process is a nonlinear version of the excitation of magnonic leaky-modes described in \cite{sobucki23nano}. Here the excitation of leaky-modes is facilitated by three consecutive nonlinear processes which finally create a wide range of SW plane waves in the system. Moreover, the newly excited SW plane waves are shifted in phase compared to reflected and scattered SWs, what is confirmed by the presence of a destructive interference pattern next to the SW beams. 
    The results presented in this paper constitute a new bridge between magnonics and nonlinear optics contributing to a new subfield of magnonics called nonlinear SW optics.

    \section*{\label{sec:met} Methods}






    \subsection*{Micromagnetic simulations}
    
    In our research we employ micromagnetic simulations performed in Mumax3 environment \cite{vansteenkiste2014design} to solve Landau-Lifshitz equation in time domain. The  system is modeled as a cuboidal layer of Py with dimensions $5.12 \mathrm{\mu m} \times Y \mathrm{\mu m} \times 10 \mathrm{nm}$ (along $x$, $y$, $z$ axes respectively). The value $Y$ varies in simulations with different incident SW beam's angle of propagation, namely  $Y =\{10.24 \mathrm{\mu m}$, $12.42 \mathrm{\mu m}$, $14.88 \mathrm{\mu m}$, $17.73 \mathrm{\mu m}\}$ for angles $\theta=\{30^\circ, 35^\circ, 40^\circ, 45^\circ\}$ respectively.
    The material parameters of Py used in the simulations are $\alpha = 0.0001$, $\mathrm{M_S} = 800$~kA/m, $\mathrm{A_{ex}}=13$~pJ/m, which yield the exchange length of $\lambda_\mathrm{ex}=5.69$~nm. We use the discretization grid $5 \mathrm{nm} \times 5 \mathrm{nm} \times 10 \mathrm{nm}$ (along $x$, $y$, $z$ axes respectively) which is shorter than $\lambda_\mathrm{ex}$ in the in-plane coordinates of the layer. We place the Py layer in a uniform external magnetic field $\mathrm{B_0}=300$~mT directed in-plane, along the $y$~axis. To simulate the infinitely long system along the $x$~axis and negative direction of $y$~axis we increase value of the damping parameter $\alpha$ up to value $0.5$ parabolically over the width of $600$~nm to prevent reflections at these edges.

    The dispersion relation presented in Fig.~\ref{fig:geo}(b) was obtained as described in \cite{gruszecki2021}. A small, two-dimensional antenna was placed at the edge of the system. The antenna introduced a locally oscillating external magnetic field with time and space distribution described by $sinc$ functions with cut-off parameters $f_\mathrm{cut}=100$~GHz and $k_\mathrm{cut}=300$~rad/$\mu$m. This magnetic field excites omnidirectional SWs in the system thus, the SWs propagate both in the bulk of the system and in the demagnetizing magnetic field dip in at the edge. For the dispersion relation calculation $1000$ snapshots of the magnetization configuration were saved with time step of $0.5/(1.1f_\mathrm{cut})$.

    To excite the SWs in the simulations with beams scattering we use two antennas, one placed in the bulk of the Py layer and the second at the system's edge. The first antenna creates SW beam aimed at the edge. We used a mathematical formula to create this antenna in the simulations inspired by Ref.\cite{whitehead2019graded}
        \begin{equation}
            \begin{split}
                B_{\mathrm{ext}, y}(t,x',y') = A(1-e^{-0.2 \pi f t})R(y')G(x') \\
                \times [  \mathrm{sin}(k y')\mathrm{sin}(2 \pi f t) + \mathrm{cos}(k y')\mathrm{cos}(2 \pi f t) ],
            \end{split}
            \label{eq:sw}
        \end{equation}
    where $A=0.01B_0$ is the amplitude of the dynamic field, $R(y')=\Theta(-y'+\frac{w_a}{2})\Theta(y'+\frac{w_a}{2})$ is a rectangle function, which describes antenna's shape along its $y^\prime$ coordinate ($\Theta$ is Heaviside step function, antenna's width $w_a=360$~nm), $G(x')=\mathrm{exp}(-\frac{x'^2}{4\sigma_x^{2}})$ is a Gaussian function defining antenna's shape along the $x'$-axis ($\sigma_x=320$~nm), $k$ is the wavevector and $f$~GHz is the frequency of the excited SWs.
    Such a formula excites in the simulations a unidirectional SW beam of Gaussian envelope with FWHM$=760$~nm. 
    The incident SW beam in all simulations has frequency $f=45$~GHz and corresponding wavevector for this SW beam is derived from Kalinikos-Slavin formula for SW dispersion relation \cite{kalinikos1986theory} for each incident SW beam's angle of propagation.
    We control the angle of the incident SW beam propagation (angle of the wave vector) by changing the rotation of the antenna, $\theta$, with respect to the edge of the system. In these investigations, we limit ourselves to the range of $\theta \in \langle 30^\circ, 45^\circ \rangle$. However, the calculated angles of propagation of the incident SW beams are slightly different because of the SW anisotropy of propagation. The second antenna is defined as a point source of Gaussian shape ($\sigma_\mathrm{edge}=15$~nm) placed at the very edge of the system. The purpose of this antenna is to excite the localized edge modes characterized by frequencies in range $\nu \in \langle 11, 15.5 \rangle$~GHz. Depending on the desired wavevector of the edge mode $\kappa$ we place this antenna either below the point where the incident SW beam reaches the edge (positive $\kappa$) or above this point (negative $\kappa$).

    The numerical simulations consist of three phases. In the first we obtain a stable static magnetic configuration of the defined external magnetic field by minimizing the energy of the initial magnetic state. In the next phase we run the dynamical part of simulation when both antennas are turned on until the system reaches the steady-state. The simulation time varies with the system's length along its $x$-axis, $t =\{240 \mathrm{ns}$, $300 \mathrm{ns}$, $350 \mathrm{ns}$, $420 \mathrm{ns}\}$ for the incident SW beam angles $\theta=\{30^\circ, 35^\circ, 40^\circ, 45^\circ\}$ respectively. Finally, we save $800$ snapshots of the magnetic configuration with $\mathrm{dt}=5$~ps time step.

    \subsection*{Simulation data postprocessing}

    To process the data obtained in the micromagnetic simulations we use a self-developed code.   
    We start with the spectral analysis of the scattered SWs by calculating Fourier transform in time domain using saved magnetic configuration snapshots for each simulation case. From this analysis we obtain an insight into the frequencies of the processes that undergo in our system and the distribution of complex SW amplitude at a given frequency ($m_z(f)\in \textbf{C}$), therefore, allowing us to analyze both the amplitude and phase of SWs in each point of the simulated area. The time sampling was chosen to provide such resolution in frequency domain to analyze the system's response in all anticipated frequencies of $f\pm n\nu$, $n \in \textbf{N}^{+}$ configurations, namely $\mathrm{dt}=5$~ps time step.
    In our calculations we investigate following frequencies $f=45$~GHz corresponding to the incident and reflected SW beams, $f-\nu$ corresponding to scattered beam in SSP, and $f+\nu$ corresponding to scattered beam in CP. Even though higher order nonlinear processes are also present in the spectral analysis we omit them as they are beyond the scope of this paper.

    \begin{figure}[!h]
        \centering
        \includegraphics[width=0.99\columnwidth]{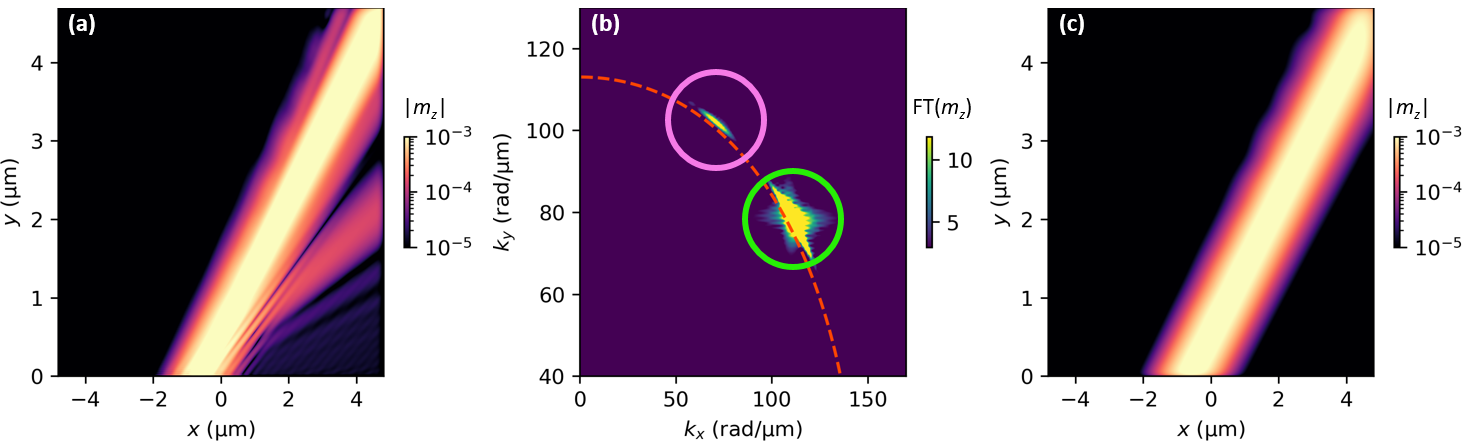}     
        \caption{ \textbf{Numerical wavevector filter.}
        (a) The $|m_z|$ amplitude space distribution of scattered SWs in SSP for $\theta=30^{\circ}$, $\kappa > 0$ and $\nu = 11$~GHz case with visible waves that are the result of secondary scattering on reflected edge mode. (b) Fourier transform of the scattered waves presented in (a). Two distinctive peaks are evident which one marked with violet circle corresponds to SSP and the second, green circle, corresponds to secondary scattering. (c) The simulation results after applying the wavevector filtering where only contribution from the wavevector of SSP is taken into consideration.
     }
        \label{fig:filter}
    \end{figure}

    The simulation results for $\kappa > 0$ and low edge mode frequency $\nu=11$~GHz proved to be problematic in further postprocessing. The reason for that was unexpected reflection of low-frequency edge mode from the absorbing boundary conditions, where up to $10~\%$ of edge mode amplitude was reflected. It caused a secondary scattering in the system and creation of an additional SW beam of frequency corresponding to SSP but propagating under different angle, as shown in Fig.~\ref{fig:filter}(a) $|m_z|$ amplitude for SSP in $\theta=30^{\circ}$, $\kappa > 0$ and $\nu = 11$~GHz case. These new beam is actually a result of a CP with negative values of edge mode frequency and wavevector, as the reflected edge mode propagates in opposite direction to the excited edge mode, thus resembling SSP in frequency but differs in wavevector of the scattered beam. To limit the effect of secondary scattering in our analysis we use an additional numerical filter which cuts off the contribution of SWs with different wavevectors than predicted by the investigated inelastic scattering processes. For that we only take the contribution from the peak and its vicinity in spectral analysis that correspond to the investigated process, as shown with violet circle in Fig.~\ref{fig:filter}(b). To extract only the contribution for the desired peak and avoid numerical errors in the further steps of postprocessing we multiply the amplitude distribution by an amplitude mask in the shape of two-dimensional Gaussian curve centered over the peak with a small spread.     
    Then the inverse Fourier transform  of the product of amplitude distribution and mask yields the SW amplitude space distribution without any undesired distortions in the system, as shown in Fig.~\ref{fig:filter}(c). This procedure was applied to the simulation results with low edge mode frequency, $\nu=11$~GHz to increase the precision of derivation of beams' trajectories.

    \subsection*{Derivation of SW beams trajectories}

    To derive the trajectories of the investigated SW beams we use distributions of SW intensity in space for given frequency and wavevector corresponding to incidence, reflection, SSP, and CP. We calculate all of the beams' trajectories in the far-field, in terms of geometry used in the simulations, $2\mathrm{\mu m}$ away from the edge. We fit Gaussian function to the cross-sections of the SW $|m_z|$ intensity distributions at fixed $y$-coordinate. We used $200$ cross-sections with an interval $5$~nm, as the discretization grid used in the simulations. After fitting the Gaussian function to each cutline we save the position of the curve's center of all  the beams. Later, we use these coordinates to fit line functions to them that we interpret as beams' trajectories. These trajectories are extended from the far field to the system's edge where the position of beams' interception with the edge are determined. To calculate the spatial shifts of the scattered beams we take as a point of reference the position where  incident beam reaches system's edge. Thus the negative value of spatial shifts means that the given beam has its origin left to the incidence spot and positive value means that the beam shifts to the right of the incidence spot.

\begin{acknowledgement}

The research leading to these results has received funding from the Polish National Science Centre projects No. 2019/35/D/ST3/03729 and 2022/45/N/ST3/01844. The numerical simulations were performed at the Poznan Supercomputing and Networking Center (Grant No. 398). Krzysztof Sobucki is a scholarship recipient of the Adam Mickiewicz University Foundation for the academic year 2023/2024.

\end{acknowledgement}

\begin{suppinfo}

 Correspondence and requests for materials should be addressed to K.S., the email address provided with affiliations.

\end{suppinfo}

\bibliography{achemso-demo}

\end{document}